\documentclass[aps,onecolumn,12pt]{revtex4}

\newcommand{\be}{\begin{equation}}
\newcommand{\ee}{\end{equation}}

\usepackage{epsfig,amsmath,amssymb} 
\usepackage{graphics}

\begin{document}

\title{Electromagnetic field of fractal distribution of charged particles}

\author{Vasily E. Tarasov}

\affiliation{\it Skobeltsyn Institute of Nuclear Physics,
Moscow State University, Moscow 119992, Russia }

\email{tarasov@theory.sinp.msu.ru}

\begin{abstract}
Electric and magnetic fields 
of fractal distribution of charged particles are considered.
The fractional integrals are used to describe fractal distribution.
The fractional integrals are considered as approximations of
integrals on fractals.  
Using the fractional generalization of integral Maxwell equation, 
the simple examples of the fields of homogeneous
fractal distribution are considered. 
The electric dipole and quadrupole moments
for fractal distribution are derived. 
\end{abstract}

%\pacs{03.50.De; 05.45.Df; 41.20.-q}
%%%PACS:  03.50.De; 05.45.Df; 41.20.-q  \\

%05.45.Df Fractals
%47.53.+n Fractals in fluid dynamics
%03.50.De Classical electromagnetism, Maxwell equations 
%41.20.-q Applied classical electromagnetism 
%41.20.Cv Electrostatics; Poisson and Laplace equations, 
%41.20.Gz Magnetostatics; magnetic shielding, magnetic induction,
%41.75.-i Charged-particle beams

%\keywords{Fractional integrals, Fractal distribution, Maxwell equations}

\maketitle

%%%\noindent
%%%PACS:  03.50.De; 05.45.Df; 41.20.-q  \\
%%%Keywords: Fractional integrals, Fractal distribution, Maxwell equations\\

\section{Introduction}

Derivatives and integrals of fractional order \cite{SKM} have found 
many applications in recent studies in physics.
The interest in fractional analysis has been growing continually 
in recent years.
Fractional analysis has numerous applications: 
kinetic theories \cite{Zaslavsky1,Zaslavsky2,Physica};  
statistical mechanics \cite{chaos,PRE05,JPCS}; 
dynamics in complex media \cite{Nig,PLA05,PLA05-2,AP05,Chaos05}; 
electromagnetic theory \cite{En1,En2,En3,En4} and many others. 
The new type of problem has increased rapidly in areas 
in which the fractal features of a process or the medium 
impose the necessity of using nontraditional tools 
in "regular" smooth physical equations. 
In order to use fractional derivatives and fractional integrals 
for fractal distribution, we must use some 
continuous medium model \cite{PLA05}. 
We propose to  describe the fractal distribution by a fractional 
continuous medium \cite{PLA05}, where all characteristics and 
fields are defined 
everywhere in the volume but they follow some generalized 
equations which are derived by using fractional integrals.
In many problems the real fractal structure of medium
can be disregarded and the fractal distribution can be replaced by
some fractional continuous mathematical model.
Smoothing of microscopic characteristics over the
physically infinitesimal volume transforms the initial
fractal distribution into fractional continuous model \cite{PLA05}
that uses the fractional integrals.
The order of fractional integral is equal
to the fractal dimension of distribution. 
The fractional integrals allow us to take into account the
fractality of the distribution.
Fractional integrals are considered as approximations 
of integrals on fractals \cite{RLWQ,Nig4}. 
In Ref. \cite{RLWQ}, the authors proved that integrals
on net of fractals can be approximated by fractional integrals.
In Ref. \cite{chaos}, we proved that fractional integrals
can be considered as integrals over the space with fractional
dimension up to numerical factor. In order to prove, we use the 
formulas of dimensional regularizations \cite{Col}.

We can consider electric and magnetic fields 
of fractal distribution of charged particles.
Fractal distribution can be described by fractional continuous medium model
\cite{PLA05,AP05,Physica,Chaos05}. 
In the general case, the fractal distribution of particles cannot 
be considered as continuous medium.
There are points and domains that have no particles.
In Ref. \cite{PLA05}, we suggest to consider the fractal distributions  
as special (fractional) continuous media.
We use the procedure of replacement of the distribution 
with fractal mass dimension by some continuous model that 
uses fractional integrals. This procedure is a fractional 
generalization of Christensen approach \cite{Chr}.
Suggested procedure leads to the fractional integration and 
differentiation to describe fractal distribution.
In this paper, we consider the electric and magnetic fields 
of fractal distribution of charged particles.
In Sec. II, the densities of electric charge and current 
for fractal distribution  are considered. 
In Sec. III and Sec. IV, we consider the simple examples of the fields of 
homogeneous fractal distribution. 
In Sec. V, we consider the fractional generalization of 
integral Maxwell equation. In Sec. VI, the examples of 
electric dipole and quadrupole moments
for fractal distribution are considered. 
Finally, a short conclusion is given in Sec. VII.

\section{Electric charge and current densities}

\subsection{Electric charge for fractal distribution}

Let us consider a fractal distribution of charged particles.
For example, we can assume that charged particles 
with a constant density are distributed over the fractal. 
In this case, the number of particles $N$ enclosed 
in a volume with characteristic size $R$ satisfies the 
scaling law $N(R) \sim R^{D}$,
whereas for a regular n-dimensional Euclidean object 
we have $N(R)\sim R^n$. 

For charged particles with number density $n({\bf r},t)$,
we have that the charge density can be defined by
\[ \rho({\bf r},t)=q n({\bf r},t) , \]
where $q$ is the charge of a particle (for electron, $q=-e$). 
The total charge of region $W$ is then given by the integral
\[ Q(W)=\int_W \rho({\bf r},t) dV_3 , \]
or $Q(W)=qN(W)$, where $N(W)$ is a number of particles in the region $W$. 
The fractional generalization of this equation can be written
in the following form
\[ Q(W)=\int_W \rho({\bf r},t) dV_D , \]
where $D$ is a fractal dimension of the distribution,
and $dV_D$ is an element of $D$-dimensional volume such that
\be \label{5a} dV_D=c_3(D,{\bf r})dV_3. \ee

For the Riesz definition of the fractional integral \cite{SKM}, the 
function $c_3(D,{\bf r})$ is defined by the relation
\be \label{5R} c_3(D,{\bf r})=
\frac{2^{3-D}\Gamma(3/2)}{\Gamma(D/2)} |{\bf r}|^{D-3} . \ee
The initial points of the fractional integral are set to zero.
The numerical factor in Eq. (\ref{5R}) has this form in order to
derive usual integral in the limit $D\rightarrow (3-0)$.
Note that the usual numerical factor
$\gamma^{-1}_3(D)={\Gamma(1/2)}/{[2^D \pi^{3/2} \Gamma(D/2)]}$,
which is used in Ref. \cite{SKM}
leads to $\gamma^{-1}_3(3-0)= {\Gamma(1/2)}/{[2^3 \pi^{3/2}
\Gamma(3/2)]}$ in the limit $D\rightarrow (3-0)$.

For the Riemann-Liouville fractional integral \cite{SKM}, 
the function $c_3(D,{\bf r})$ is defined by
\be \label{5RL} c_3(D,{\bf r})=
\frac{|x y z |^{D/3-1}}{\Gamma^3(D/3)}  . \ee
Here we use Cartesian's coordinates $x$, $y$, and $z$. 
In order to have the usual dimensions of the physical values,
we can use vector ${\bf r}$, and coordinates 
$x$, $y$, $z$ as dimensionless values.

Note that the interpretation of fractional integration
is connected with fractional dimension \cite{chaos,PRE05}.
This interpretation follows from the well-known formulas 
for dimensional regularizations \cite{Col}.
The fractional integral can be considered as an
integral in the fractional dimension space
up to the numerical factor $\Gamma(D/2) / [ 2 \pi^{D/2} \Gamma(D)]$.

If we consider the ball region $W=\{{\bf r}: \ |{\bf r}|\le R \}$, 
and the spherically symmetric distribution of charged particles 
($n({\bf r},t)=n(r)$), then we have
\[ N(R)=4\pi \frac{2^{3-D}\Gamma(3/2)}{\Gamma(D/2)}
\int^R_0 n(r) r^{D-1} dr , \quad Q(R)=qN(R). \]
For the homogeneous ($n(r)=n_0$) fractal distribution, we get
\[ N(R)=4\pi n_0 \frac{2^{3-D}\Gamma(3/2)}{\Gamma(D/2)}
\frac{R^D}{D} \sim R^D . \]
Fractal distribution of particles is called a homogeneous 
fractal distribution if the power law  $N(R)\sim R^D $ does not 
depend on the translation  of the region. 
The homogeneity property of the distribution
can be formulated in the following form:
For all regions, $W$ and $W^{\prime}$ such that 
the volumes are equal, $V(W)=V(W^{\prime})$, 
we have that the numbers of particles in these regions 
are equal too, $N(W)=N(W^{\prime})$. 
Note that the wide class of fractal media satisfies 
the homogeneous property.
In Ref. \cite{PLA05}, the continuous medium model for the fractal 
distribution of particles was suggested. Note that the fractality and 
homogeneity properties can be realized in the following forms: 

\noindent
(1) Homogeneity:
The local number density of homogeneous fractal distribution 
is a translation invariant value that has the form
$n({\bf r})=n_0=const$.

\noindent
(2) Fractality:
The number of particles in the ball region $W$ obeys a power law relation
$N_D(W) \sim R^D$, where $D<3$, $R$ is the radius of the ball.

%%%%%%%%%%%%%%%%%%%%%%%%%%%%%%%%%%%%%%%%%%%%%%%%%%%%%%%%%%%%%%%%%%
\subsection{Electric current of fractal distribution}

For charged particles with number density $n({\bf r},t)$ flowing 
with velocity ${\bf u}={\bf u}({\bf r},t)$, 
the resulting density current ${\bf J}({\bf r},t)$ is given by
\[ {\bf J}({\bf r},t)= q n({\bf r},t) {\bf u} , \]
where $q$ is the charge of a particle (for electron, $q=-e$). 

The electric current is defined as the flux of electric charge.
Measuring the field ${\bf J}({\bf r},t)$ passing through a surface 
$S=\partial W$ gives the current (flux of charge) 
\[ I(S)=\Phi_J(S)=\int_S ({\bf J}, d{\bf S}_2) , \]
where ${\bf J}={\bf J}({\bf r},t)$ is the current field vector, 
$d{\bf S_2}=dS_2{\bf n}$ is a differential unit of area 
pointing perpendicular to the surface $S$, 
and the vector ${\bf n}=n_k {\bf e}_k$ is a vector of normal.
The fractional generalization of this equation for the fractal 
distribution can be written in the following form:
\[ I(S)=\int_S ({\bf J}({\bf r},t), d{\bf S}_d) , \]
where we use
\be \label{C2} dS_d=c_2 (d,{\bf r})dS_2 , \quad 
c_2(d,{\bf r})= \frac{2^{2-d}}{\Gamma(d/2)} |{\bf r}|^{d-2} . \ee
Note that $c_2(2,{\bf r})=1$ for $d=2$. 
In general, the boundary $\partial W$ has the dimension $d$. 
In the general case, the dimension $d$ is not equal to $2$ and 
is not equal to $(D-1)$.

\subsection{Charge conservation for fractal distribution}

The electric charge has a fundamental property established
by numerous experiments: the change of the quantity of charge
inside a region $W$ bounded by the surface $S=\partial W$
is always equal to the flux of charge through this surface.
This is known as the law  of charge conservation. If we denote
by ${\bf J}({\bf r},t)$ the electric current density,
then charge conservation is written
\[ \frac{dQ(W)}{dt}=-I(S), \]
or, in the form
\be \label{cecl} \frac{d}{dt} \int_W \rho({\bf r},t) dV_D= 
- \oint_{\partial W} ({\bf J} ({\bf r},t),d{\bf S}_d) . \ee
In particular, when the surface $S=\partial W$ is fixed,
we can write
\be \label{drho} \frac{d}{dt} \int_W \rho({\bf r},t) dV_D= 
\int_W \frac{\partial \rho({\bf r},t)}{\partial t} dV_D .\ee
Using the fractional generalization of the 
mathematical Gauss's theorem (see Appendix), we have
\be \label{gt} \oint_{\partial W} ({\bf J} ({\bf r},t),d{\bf S}_d) =
\int_W c^{-1}_3(D,{\bf r})
\frac{\partial}{\partial x_k} \Bigl( c_2(d,{\bf r})J_k({\bf r},t) \Bigr)
dV_D .\ee
Substituting the right hand sides of Eqs. (\ref{drho}) and (\ref{gt})
in Eq. (\ref{cecl}), we find the law of charge 
conservation in differential form
\[ c_3(D,{\bf r})\frac{\partial \rho({\bf r},t)}{\partial t}+
\frac{\partial}{\partial x_k} \Bigl( c_2(d,{\bf r})J_k({\bf r},t) \Bigr) =0 . \]
This equation can be considered as a continuity equation for
fractal distribution of particles \cite{AP05}.

\section{Electric field of fractal distribution}

\subsection{Electric field and Coulomb's law}

For a point charge $Q$ at position  ${\bf r}^{\prime}$ 
(i.e., an electric monopole), 
the electric field at a point ${\bf r}$ is defined in MKS by
\[ {\bf E}=\frac{Q}{4 \pi \varepsilon_0} \
\frac{{\bf r}-{\bf r}^{\prime}}{|{\bf r}-{\bf r}^{\prime}|^3} , \]
where $\varepsilon_0$
%%%=8.8542 \cdot 10^{-12} Fm^{-1}$ 
is a fundamental constant called the permittivity of free space. 

For a continuous stationary distribution $\rho({\bf r}^{\prime})$ of charge, 
the electric field ${\bf E}$ at a point ${\bf r}$ is given in MKS by
\be \label{E}
{\bf E}({\bf r})=\frac{1}{4 \pi \varepsilon_0} \int_W
\frac{{\bf r}-{\bf r}^{\prime}}{|{\bf r}-{\bf r}^{\prime}|^3}
\rho({\bf r}^{\prime}) dV^{\prime}_3 , 
\ee
where $\varepsilon_0$ is the permittivity of free space. 
For Cartesian's coordinates $dV^{\prime}_3=dx^{\prime}dy^{\prime}dz^{\prime}$.

The fractional generalization of Eq. (\ref{E}) for 
a fractal distribution of charge is given by the equation
\be \label{CLD}
{\bf E}({\bf r})=\frac{1}{4 \pi \varepsilon_0} \int_W
\frac{{\bf r}-{\bf r}^{\prime}}{|{\bf r}-{\bf r}^{\prime}|^3}
\rho({\bf r}^{\prime}) dV^{\prime}_D , \ee
where $dV^{\prime}_D=c_3(D,{\bf r}^{\prime}) dV^{\prime}_3$. 
Equation (\ref{CLD}) can be considered as Coulomb's law written 
for a fractal stationary distribution of electric charges. 

Measuring the electric field passing through a surface 
$S=\partial W$ gives the electric flux 
\[ \Phi_E(S)=\int_S ({\bf E}, d{\bf S}_2) , \]
where ${\bf E}$ is the electric field vector, and $d{\bf S}_2$ 
is a differential unit of area pointing perpendicular to the surface S.

\subsection{Gauss's law for fractal distribution}

Gauss's law tells us that the total flux $\Phi_E(S)$ of 
the electric field ${\bf E}$
through a closed surface $S=\partial W$ 
is proportional to the total electric charge $Q(W)$
inside the surface: 
\be \label{GL1} \Phi_E(\partial W)=\frac{1}{\varepsilon_0} Q(W) . \ee
The electric flux out of any closed surface 
is proportional to the total charge enclosed within the surface. 

For the fractal distribution, Gauss's law states
\be \label{GL2} \int_S ({\bf E},d{\bf S}_2)=\frac{1}{\varepsilon_0} 
\int_W \rho ({\bf r},t) dV_D \ee
in MKS, 
where ${\bf E}={\bf E}({\bf r},t)$ is the electric field, and 
$\rho({\bf r},t)$ is the 
charge density,  $dV_D=c_3(D,{\bf r})dV_3$,
and $\varepsilon_0$ is the permittivity of free space. 

Gauss's law  by itself can be used to find the electric field 
of a point charge at rest, and the principle of superposition 
can then be used to find the electric field of an arbitrary 
fractal charge distribution.

If we consider the stationary spherically symmetric fractal distribution
$\rho({\bf r},t)=\rho(r)$, and the ball region 
$W=\{{\bf r}:\ |{\bf r}|\le R\}$, then we have
\[ Q(W)=4 \pi \int^R_0 \rho(r) c_3(D,{\bf r}) r^2 dr , \]
where $c_3(D,{\bf r})$ is defined in Eq. (\ref{5R}), i.e., 
\be \label{QW} Q(W)=4 \pi \frac{2^{3-D}\Gamma(3/2)}{\Gamma(D/2)}
\int^R_0 \rho(r) r^{D-1} dr . \ee
Using the sphere $S=\{{\bf r}: \ |{\bf r}|= R \}$ as a 
surface $S=\partial W$, we get
\be \label{PW} \Phi_E(\partial W)= 4 \pi R^2 E(R). \ee
Substituting Eqs. (\ref{QW}) and (\ref{PW}) in Gauss's law (\ref{GL1}),
we get the equation for electric field.
As a result, Gauss's law for fractal distribution 
with spherical symmetry leads us to the equation for electric field
\[ E(R)=\frac{2^{3-D}\Gamma(3/2)}{\varepsilon_0 R^2 \Gamma(D/2)}
\int^R_0 \rho(r) r^{D-1} dr .\]
For example, the electric field of homogeneous ($\rho(\bf r)=\rho$)
spherically symmetric fractal distribution is defined by
\[ E(R)=\rho \frac{2^{3-D}\Gamma(3/2)}{\varepsilon_0 D \Gamma(D/2)} 
R^{D-2}  \sim R^{D-2} .\]

\section{Magnetic field of fractal distribution}

\subsection{Magnetic field and Biot-Savart law}
    
The Biot-Savart law relates magnetic fields to the currents 
which are their sources. 
In a similar manner, Coulomb's law relates electric fields 
to the point charges which are their sources. 
Finding the magnetic field resulting from a fractal current distribution 
involves the vector product and is inherently a fractional calculus 
problem when the distance from the current to the field point 
is continuously changing. 

For a continuous distribution the Biot-Savart law in MKS has the form
\be \label{BSL0} {\bf B}({\bf r})=\frac{\mu_0}{4\pi} \int_W 
\frac{[{\bf J}({\bf r}^{\prime}),{\bf r}-{\bf r}^{\prime}]}{
|{\bf r}-{\bf r}^{\prime}|^3} d V^{\prime}_3 , \ee
where $[\ , \ ]$ is a vector product, 
${\bf J}$ is the current density, 
$\mu_0$ is the permeability of free space. 

The fractional generalization of Eq. (\ref{BSL0}) for
a fractal distribution in MKS has the form
\be \label{BSL} {\bf B}({\bf r})=\frac{\mu_0}{4\pi} \int_W 
\frac{[{\bf J}({\bf r}^{\prime}),{\bf r}-{\bf r}^{\prime}]}{
|{\bf r}-{\bf r}^{\prime}|^3} d V^{\prime}_D . \ee
This equation can be considered as Biot-Savart law written 
for a steady current with fractal distribution of electric charges.
The Biot-Savart law (\ref{BSL}) can be used to find the magnetic 
field produced by any fractal distribution of steady currents.

\subsection{Ampere's law for fractal distribution}

The magnetic field in space around an electric current 
is proportional to the electric current which serves as its source, 
just as the electric field in space is proportional to 
the charge which serves as its source. 
In the case of static electric field, 
the line integral of the magnetic field around 
a closed loop is proportional to the electric current 
flowing through the loop. The Ampere's law 
is equivalent to the steady state of the integral Maxwell equation 
in free space, and relates the spatially varying magnetic field 
${\bf B}({\bf r})$ to the current density ${\bf J}({\bf r})$. 

Note that, as mentioned by Lutzen in his article \cite{Lutzen}, 
Liouville, who was one of pioneers in development of 
fractional  calculus, was inspired by the problem of 
fundamental force law in Ampbre's electrodynamics and 
used fractional differential equation in that  problem.  

Let be a closed path around a current. Ampere's law states that
the line integral of the magnetic field ${\bf B}$ along 
the closed path $L$ is given in MKS by
\[ \oint_L ({\bf B},d{\bf l})=\mu_0 I(S) , \]
where $d{\bf l}$ is the differential length element, and
$\mu_0$ is the permeability of free space. 
For the fractal distribution of charged particles, we use
\[ I(S)=\int_S ({\bf J},d{\bf S}_d) , \]
where $d {\bf S}_d=c_2(d,{\bf r}) d S_2$.

If we consider the cylindrically symmetric fractal distribution, 
we have 
\[ I(S)=2 \pi \int^R_0 J(r) c_2(d,{\bf r}) r dr , \]
where $c_2(d,{\bf r})$ is defined in Eq. (\ref{C2}), i.e., 
\[ I(S)=4 \pi \frac{2^{2-d}}{\Gamma(d/2)}
\int^R_0 J(r) r^{d-1} dr . \]
Using the circle $L=\partial W=\{{\bf r}: \ |{\bf r}|=R \}$, we get
\[ \oint_L ({\bf B},d{\bf l})= 2 \pi R \ B(R). \]
As a result, Ampere's law for fractal distribution with 
cylindrical symmetry leads us to the equation for magnetic field
\[ B(R)= \frac{ \mu_0 2^{2-d}}{R \Gamma(d/2)}
\int^R_0 J(r) r^{d-1} dr .\]
For example, the magnetic field $B(r)$ of homogeneous 
($J(r)=J_0$) fractal distribution is defined by
\[ B(R)=J_0 \frac{\mu_0 2^{2-d}}{d \Gamma(d/2)} R^{d-1} \sim R^{d-1} .\]

\section{Fractional integral Maxwell equations} 
    
The Maxwell equations are the set of 
fundamental equations governing electromagnetism 
(i.e., the behavior of electric and magnetic fields).
The equations that can be expressed in integral form
are known as Gauss's law, Faraday's law, 
the absence of magnetic monopoles, and Ampere's law 
with displacement current.
In MKS,  these become 
\[ \oint_S ({\bf E},d{\bf S}_2)=
\frac{1}{\varepsilon_0} \int_W \rho dV_D , \]
\[ \oint_L ({\bf E},d{\bf l}_1)=
-\frac{\partial}{\partial t} \int_S ({\bf B},d{\bf S}_2) , \]
\[ \oint_S ({\bf B},d{\bf S}_2)= 0, \]
\[ \oint_L ({\bf B},d{\bf l}_1)=\mu_0 \int_S ({\bf J}, d{\bf S}_d)
+ \varepsilon_0 \mu_0\frac{\partial}{\partial t}
\int_S ({\bf E},d{\bf S}_2) . \]

Let us consider the special case such that 
the fields are defined on fractal \cite{Feder} only.
The hydrodynamic and thermodynamics fields 
can be defined in the fractal media \cite{AP05,Physica}.
Suppose that the electromagnetic field can be defined on fractal 
as an approximation of some real case with fractal medium.
If the electric field ${\bf E}({\bf r})$ and magnetic 
field ${\bf B}({\bf r})$ can be defined on fractal and 
does not exist outside of fractal in Eucledian space $E^3$, 
then we must use the fractional generalization of the
integral Maxwell equations in the form
\[ \oint_S ({\bf E},d{\bf S}_d)=
\frac{1}{\varepsilon_0} \int_W \rho dV_D , \]
\[ \oint_L ({\bf E},d{\bf l}_{\gamma})=
-\frac{\partial}{\partial t} \int_S ({\bf B},d{\bf S}_d) , \]
\[ \oint_S ({\bf B},d{\bf S}_d)= 0, \]
\[ \oint_L ({\bf B},d{\bf l_{\gamma}})=\mu_0 \int_S ({\bf J}, d{\bf S}_d)
+ \varepsilon_0 \mu_0\frac{\partial}{\partial t}
\int_S ({\bf E},d{\bf S}_d) . \]
These fractional integral equations have unusual properties.
Note that fractional integrals are considered as an approximation 
of integrals on fractals \cite{RLWQ,Nig4}. 

Using the fractional generalization of Stokes's and Gauss's
theorems (see Appendix), 
we can rewrite the fractional integral Maxwell equations in the form
\[ \int_W c^{-1}_3(D,{\bf r}) div( c_2(d,{\bf r}) {\bf E}) dV_D 
= \frac{1}{\varepsilon_0} \int_W \rho dV_D , \]
\[ \int_S  c^{-1}_2(d,{\bf r})
( curl(c_1(\gamma,{\bf r}){\bf E}), d{\bf S}_d) = 
-\frac{\partial}{\partial t} \int_S ({\bf B},d{\bf S}_d) , \]
\[ \int_W c^{-1}_3(D,{\bf r}) div(c_2(d,{\bf r}) {\bf B}) dV_d=0, \]
\[ \int_S c^{-1}_2(d,{\bf r}) 
(curl(c_1(\gamma,{\bf r}){\bf B}), d{\bf S}_d) = 
\mu_0 \int_S ({\bf J}, d{\bf S}_d)
+ \varepsilon_0 \mu_0\frac{\partial}{\partial t}
\int_S ({\bf E},d{\bf S}_d) ,  \]
As a result, we have the following differential Maxwell equations: 
\[ div \Bigl(c_2(d,{\bf r}) {\bf E} \Bigr) = 
\frac{1}{\varepsilon_0} c_3(D,{\bf r}) \rho  , \]
\[ curl \Bigl(c_1(\gamma,{\bf r}){\bf E} \Bigr)= 
-c_2(d,{\bf r}) \frac{\partial}{\partial t} {\bf B} , \]
\[ div \Bigl( c_2(d,{\bf r}) {\bf B} \Bigr)= 0, \]
\[  curl \Bigl( c_1(\gamma,{\bf r}){\bf B} \Bigr) = 
\mu_0 c_2(d,{\bf r}) {\bf J}+\varepsilon_0 \mu_0 c_2(d,{\bf r})
\frac{\partial {\bf E}}{\partial t}. \]

Note that the law of absence of magnetic monopoles 
for the fractal leads us to the equation 
$div ( c_2(d,{\bf r}) {\bf B} )= 0 $. 
This equation can be rewritten in the form
\[ div{\bf B}=-({\bf B}, grad c_2(d,{\bf r}) ) . \]
In the general case ($d \not=2$), the vector
$grad \ (c_2(d,{\bf r}))$ is not equal 
to zero and the magnetic field satisfies $div {\bf B}\not=0$.
If $d=2$, we have $div ({\bf F})\not=0$ only for nonsolenoidal 
field ${\bf F}$. Therefore the magnetic field on the fractal  
is similar to the nonsolenoidal field. 
As a result, the magnetic field on fractal can be considered as 
a field with some "fractional magnetic monopole" \
$q_m\sim ({\bf B},\nabla c_2)$.

%%%%%%%%%%%%%%%%%%%%%%%%%%%%%%%%%%%%%%%%%%%%%%%%%%%%%%%%%%%%%
\section{Multipole moments for fractal distribution}

\subsection{Electric multipole expansion}

A multipole expansion is a series expansion of the effect produced 
by a given system in terms of an expansion parameter which becomes 
small as the distance away from the system increases. 
Therefore, the leading one of the terms in a multipole expansion are 
generally the strongest. The first-order behavior of the system 
at large distances can therefore be obtained from the first terms 
of this series, which is generally much easier to compute than 
the general solution. Multipole expansions are most commonly used 
in problems involving the gravitational field of mass aggregations, 
the electric and magnetic fields of charge and current distributions, 
and the propagation of electromagnetic waves.

To compute one particular case of a multipole expansion, 
let ${\bf R}=X_k{\bf e}_k$ be the vector from a fixed reference point to 
the observation point, ${\bf r}=x_k{\bf e}_k$ 
be the vector from the reference 
point to a point in the body, and ${\bf d}={\bf R}-{\bf r}$
be the vector from a point in the body to the observation point. 
The law of cosines then yields
\[ d^2=R^2+r^2-2rR \cos \theta =
R^2\Bigl( 1+\frac{r^2}{R^2} -2\frac{r}{R} \cos \theta \Bigr) , \]
where $d=|{\bf d}|$, and $\cos \theta = ({\bf r},{\bf R})/(r R)$, so
\[ d=R \sqrt{ 1+\frac{r^2}{R^2} -2\frac{r}{R} \cos \theta } . \]
Now define $\epsilon ={r}/{R}$, and $x=\cos \theta$, then
\[ \frac{1}{d}=\frac{1}{R} 
\Bigl( 1-2 \epsilon x+\epsilon^2 \Bigr)^{-1/2}. \]

But  $\Bigl( 1-2 \epsilon x+\epsilon^2 \Bigr)^{-1/2}$
is the generating function for Legendre polynomials $P_n(x)$ as follows: 
\[ \Bigl( 1-2 \epsilon x+\epsilon^2 \Bigr)^{-1/2}=
\sum^{\infty}_{n=0} \epsilon^n P_n(x) , \]
so, we have the equation
\[ \frac{1}{d}=\frac{1}{R} \sum^{\infty}_{n=0} 
\Bigl(\frac{r}{R}\Bigr)^n P_n( \cos \theta) . \]
Any physical potential that obeys a $(1/d)$ law can therefore 
be expressed as a multipole expansion
\be \label{11}
U=  \frac{1}{4 \pi \varepsilon_0}  \sum^{\infty}_{n=0} \frac{1}{R^{n+1}}
\int_W r^n P_n( \cos \theta)  \rho({\bf r}) dV_D. 
\ee
The $n = 0$ term of this expansion, called the monopole term, 
can be pulled out by noting that $P_0(x)=1$, so
\be
U= \frac{1}{4 \pi \varepsilon_0} \frac{1}{R} \int_W \rho({\bf r}) dV_D+
\frac{1}{4 \pi \varepsilon_0}  \sum^{\infty}_{n=1} \frac{1}{R^{n+1}}
\int_W r^n P_n( \cos \theta)  \rho({\bf r}) dV_D. 
\ee
The $n$th term
\be
U_n=  \frac{1}{4 \pi \varepsilon_0}  \frac{1}{R^{n+1}}
\int_W r^n P_n( \cos \theta)  \rho({\bf r}) dV_D 
\ee
is commonly named according to the following:
n - multipole, 0 - monopole, 1 -  dipole, 2 -  quadrupole.

\subsection{Electric dipole moment of fractal distribution}

An electric multipole expansion is a determination 
of the voltage $U$ due to a collection of charges obtained 
by performing a multipole expansion. 
This corresponds to a series expansion of the charge density 
$\rho({\bf r})$ in terms of its moments, normalized by the distance 
to a point ${\bf R}$ far from the charge distribution. 
In MKS, the electric multipole expansion is given by Eq. (\ref{11}):
\be 
U=\frac{1}{4 \pi \varepsilon_0}\sum^{\infty}_{n=0} 
\frac{1}{R^{n+1}} \int_W r^n P_n(\cos \theta) \rho({\bf r}) dV_D,
\ee
where $P_n(\cos \theta)$ is a Legendre polynomial and 
$\theta$ is the polar angle, defined such that
\[ \cos \theta= ({\bf r},{\bf R})/(|{\bf r}| |{\bf R}| ) . \]

The first term arises from $P_0(x)=1$, while all further terms 
vanish as a result of $P_n(x)$ being a polynomial in $x$ for $n\ge 1$, 
giving $P_n(0)=0$ for all $n\ge 1$.

If we have
\[ Q(W)=\int_W \rho({\bf r})dV_D=0 , \]
then the $n = 0$ term vanishes. 
Set up the coordinate system so that $\theta$ measures 
the angle from the charge-charge line with the midpoint 
of this line being the origin. Then the $n = 1$ term is given by 
\[ U_1= \frac{1}{4 \pi \varepsilon_0} \frac{1}{R^2} \int_W
r P_1(\cos \theta) \rho({\bf r}) dV_D =\]
\[ =\frac{1}{4 \pi \varepsilon_0} \frac{1}{R^2} \int_W
r \ \cos \theta \rho({\bf r}) dV_D =
\frac{1}{4 \pi \varepsilon_0} \frac{1}{R^2} \int_W
\frac{({\bf r},{\bf R})}{R}  \rho({\bf r}) dV_D =\]
\[ =\frac{1}{4 \pi \varepsilon_0} \frac{1}{R^3} \int_W
 ({\bf r},{\bf R}) \rho({\bf r}) dV_D =
\frac{1}{4 \pi \varepsilon_0} \frac{1}{R^3} 
\Bigl( {\bf R},\int_W {\bf r}  \rho({\bf r}) dV_D \Bigr) .\]

For a continuous charge distribution, 
the electric dipole moment is given by 
\be \label{d1} {\bf p}=\int_W {\bf r} \rho({\bf r}) dV_3, \ee
where ${\bf r}$ points from positive to negative.
Defining the dipole moment for the fractal distribution by
the equation
\be \label{ED} {\bf p}^{(D)}=\int_W  {\bf r} \rho({\bf r})  dV_D , \ee
then gives 
\[ U_1=\frac{1}{4 \pi \varepsilon_0} 
\frac{({\bf R},{\bf p}^{(D)})}{R^3} =
\frac{1}{4 \pi \varepsilon_0} \frac{p^{(D)} \cos \alpha}{R^2}\]
where $\cos \alpha=({\bf R}, {\bf p}^{(D)})/(p^{(D)} R)$, 
and $p^{(D)}=\sqrt{(p^{(D)}_x)^2+(p^{(D)}_y)^2+(p^{(D)}_z)^2}$.

Let us consider the dipole moment for the fractal distribution by
Eq. (\ref{ED}), 
where we use the Riemann-Liouville fractional integral, and  
the function $c_3(D,{\bf r})$ in the form
\be c_3(D,{\bf r})=
\frac{|x y z |^{a-1}}{\Gamma^3(a)} , \quad a=D/3. \ee
Let us consider the example of electric dipole moment
for the homogeneous ($\rho({\bf r})=\rho$) fractal distribution
of electric charges in the parallelepiped region
\be \label{paral} 
W=\{(x;y;z): \ 0\le x \le A,\  0\le y \le B , \ 0\le z \le C \}  . \ee
In this case, we have Eq. (\ref{ED}) in the form
\[ p^{(D)}_x=\frac{\rho}{\Gamma^3(a)} 
\int^A_0 dx \int^B_0 dy \int^c_0 dz \  x^{a}y^{a-1}z^{a-1}= 
\frac{\rho (ABC)^a}{\Gamma^3(a)}  \frac{A}{a^2(a+1)} .\]
The electric charge of parallelepiped region (\ref{paral})
is defined by
\[ Q(W)=\rho \int_W dV_D=\frac{\rho (ABC)^a}{a^3 \Gamma^3(a)}  .\]
Therefore, we have the dipole moments for fractal
distribution in parallelepiped in the form
\[ p^{(D)}_x=\frac{a}{a+1} Q(W) A ,\quad 
p^{(D)}_y=\frac{a}{a+1} Q(W) B, \quad 
p^{(D)}_z=\frac{a}{a+1} Q(W) C , \]
where we can use $a/(a+1)=D/(D+3)$. 
As a result, we get
\be  p^{(D)}_k=\frac{2 D}{D+3} p^{(3)}_k , \ee
where $p^{(3)}_k$ are the dipole moments for three-dimensional 
homogeneous distribution. If we use the following limits 
$2<D\le 3$, then we have
\[ 0.8 < \frac{2 D}{D+3} \le 1 . \]

\subsection{Electric quadrupole moment of fractal distribution}

While this is the dominant term for a dipole, 
there are also higher-order terms in 
the multipole expansion that become smaller as $R$ becomes large.
The electric quadrupole term in MKS is given by 
\[ U_2= \frac{1}{4 \pi \varepsilon_0} \frac{1}{R^3} \int_W
r^2 P_2(\cos \theta) \rho({\bf r}) dV_D =\]
\[ = \frac{1}{4 \pi \varepsilon_0} \frac{1}{R^3} \int_W
r^2 \left(\frac{3}{2}\cos^2 \theta-\frac{1}{2} \right) \rho({\bf r}) dV_D =\]
\[ = \frac{1}{4 \pi \varepsilon_0} \frac{1}{2 R^3} \int_W
r^2 (3 \cos^2 \theta-1) \rho({\bf r}) dV_D =\]
\[ = \frac{1}{4 \pi \varepsilon_0} \frac{1}{2 R^3} \int_W
\left( \frac{3}{R^2}({\bf R},{\bf r})^2-r^2 \right) \rho({\bf r}) dV_D .\]

The electric quadrupole is the third term in an electric 
multipole expansion, and can be defined in MKS by
\[ U_2= \frac{1}{4 \pi \varepsilon_0}\frac{1}{2 R^3}
\sum^3_{k,l=1} \frac{X_k X_l}{R^2} Q_{kl} , \]
where $\varepsilon_0$ is the permittivity of free space, 
$R$ is the distance from the fractal distribution of charges, 
and $Q_{kl}$  is the electric quadrupole moment, which is a tensor.

The electric quadrupole moment is defined by the equation
\[ Q_{kl}=\int_W (3 x_k x_l-r^2\delta_{kl}) \rho({\bf r}) dV_D ,\]
where $x_k= x, y$, or $z$. From this definition, it follows that
\[ Q_{kl}=Q_{lk} , \quad and  \quad \sum^{3}_{k=1} Q_{kk}=0. \]
Therefore, we have $Q_{zz}=-Q_{xx}-Q_{yy}$.
In order to compute the values
\[ Q^{(D)}_{xx}=\int_W [3x^2-(x^2+y^2+z^2)] \rho({\bf r}) dV_D=
\int_W [2x^2-y^2-z^2] \rho({\bf r}) dV_D , \] 
\[ Q^{(D)}_{yy}=\int_W [3y^2-(x^2+y^2+z^2)] \rho({\bf r}) dV_D=
\int_W [-x^2+2y^2-z^2)] \rho({\bf r}) dV_D , \] 
\[ Q^{(D)}_{zz}=\int_W [3z^2-(x^2+y^2+z^2)] \rho({\bf r}) dV_D =
\int_W [-x^2-y^2+2z^2)] \rho({\bf r}) dV_D , \] 
we consider the following expression
\be \label{Qabc}
 Q(\alpha,\beta,\gamma)=
\int_W [\alpha x^2+\beta y^2+\gamma z^2)] \rho({\bf r}) dV_D ,\ee
where we use the Riemann-Liouville fractional integral \cite{SKM}, and  
the function $c_3(D,{\bf r})$ in the form
\be c_3(D,{\bf r})=\frac{|x y z |^{a-1}}{\Gamma^3(a)} , \quad a=D/3. \ee
Using Eq. (\ref{Qabc}), we have
\be \label{QQ} Q^{(D)}_{xx}=Q(2,-1,-1), \quad  Q^{(D)}_{xx}=Q(-1,2,-1),
\quad Q^{(D)}_{zz}=Q(-1,-1,2) . \ee

\subsection{Quadrupole moment of fractal parallelepiped}

Let us consider the example of electric quadrupole moment
for the homogeneous ($\rho({\bf r})=\rho$) fractal distribution
of electric charges in the parallelepiped region
\be\label{par} 
W=\{(x;y;z): \ 0 \le x \le A,\  0 \le y \le B , \ 0 \le z \le C \} . \ee
If we consider the region $W$ in form (\ref{par}), 
then we get
\[ Q(\alpha,\beta,\gamma)= \frac{\rho (ABC)^a}{(a+2)a^2 \Gamma^3(a) }
(\alpha A^2+\beta B^2+\gamma C^2) . \]
The electric charge of this region $W$ is 
\[ Q(W)=\rho \int_W dV_D=\frac{\rho (ABC)^a}{a^3 \Gamma^3(a)}  .\]
Therefore, we have the following equation
\[ Q(\alpha,\beta,\gamma)= \frac{a}{a+2} Q(W)
(\alpha A^2+\beta B^2+\gamma C^2) , \]
where $a=D/3$. If $D=3$, then we have
\[ Q(\alpha,\beta,\gamma)= \frac{1}{3} Q(W)
(\alpha A^2+\beta B^2+\gamma C^2) . \]
As a result, we get electric quadrupole moments $Q^{(D)}_{kk}$ of 
fractal distribution in the region $W$:
\[ Q^{(D)}_{kk}=\frac{3D}{D+6} \ Q^{(3)}_{kk} , \]
where $Q^{(3)}_{kk}$ are moments for the usual 
homogeneous distribution ($D=3$). 
By analogy with these equations, we can derive $Q^{(D)}_{kl}$ for the case
$k\not=l$. These electric quadrupole moments are
\[ Q^{(D)}_{kl}=\frac{4 D^2}{(D+3)^2} \ Q^{(3)}_{kl} , \quad (k\not=l). \]
If we use the following limits $2<D\le 3$, then we get the relations 
\[ 0.75 < \frac{3D}{D+6}\le 1 , \quad
0.64 < \frac{4 D^2}{(D+3)^2} \le 1 . \]

\subsection{Quadrupole moment of fractal ellipsoid}

Let us consider the example of electric quadrupole moment
for the homogeneous ($\rho({\bf r})=\rho$) fractal distribution
in the ellipsoid region $W$:
\be\label{ell} 
\frac{x^2}{A^2}+\frac{y^2}{B^2}+\frac{z^2}{C^2} \le 1 . \ee
If we consider the region $W$ in the form (\ref{ell}), then we get
(\ref{Qabc}) in the form
\[ Q(\alpha,\beta,\gamma)= \frac{\rho (ABC)^a}{ (3a+2) \Gamma^3(a) }
(\alpha A^2 K_1(a)+\beta B^2 K_2(a)+\gamma C^2 K_3 (a) ) , \]
where $a=D/3$, and $K_i(a)$ (i=1,2,3) are defined by
\[ K_1(a)=L(a+1,a-1,2\pi) L(a-1,2a+1,\pi), \]
\[ K_2(a)=L(a-1,a+1,2\pi) L(a-1,2a+1,\pi),  \]
\[ K_3(a)=L(a-1,a-1,2\pi) L(a+1,2a-1,\pi) . \]
Here we use the following function
\[ L(n,m,l)=\frac{2 l}{\pi} \int^{\pi / 2}_0 dx 
\ |\cos (x)|^{n} | \sin (x) |^m =
\frac{l}{\pi} \frac{\Gamma(n/2+1/2) \Gamma(m/2+1/2)}{\Gamma(n/2+m/2+1)}. \]
If $D=3$, we obtain
\be \label{Qabce} 
Q(\alpha,\beta,\gamma)= \frac{4\pi}{3} \frac{\rho ABC}{5}
(\alpha A^2 +\beta B^2 +\gamma C^2 ) , \ee
where we use $K_1=K_2=K_3={4\pi}/{3}$.
The total charge of this region $W$ is 
\be \label{Qe} Q(W)=\rho \int_W dV_D=
\frac{\rho (ABC)^a}{3a \Gamma^3(a)} \frac{2 \Gamma^3(a/2)}{\Gamma(3a/2)} .\ee
If $D=3$, we have the total charge
\be Q(W)=\rho \int_W dV_3=\frac{4 \pi}{3} \rho ABC . \ee

Using Eq. (\ref{Qe}), we get the electric
quadrupole moments (\ref{QQ}) for fractal ellipsoid
\be Q(\alpha,\beta,\gamma)= \frac{a}{3a+2} Q(W)
\left( \alpha A^2 +\beta B^2 +\gamma C^2 \right) , \ee
where $a=D/3$. If $D=3$, then we have the well-known relation
\[ Q(\alpha,\beta,\gamma)= \frac{Q(W)}{5}
(\alpha A^2+\beta B^2+\gamma C^2) . \]

\section{Conclusion}

In this paper, we have introduced and described  
the fractional continuous model for the 
fractal distribution of charged particles.  
Using the fractional calculus and the fractional continuous model, 
we have shown that the fractional integrals can be used
for calculation of multipole moments of the fractal distribution. 
The order of fractional integral is equal 
to the fractal dimension of the distribution.

The fractional continuous models for fractal distribution of 
particles may have applications in plasma physics. 
This is due in part to the relatively small numbers of parameters 
that define a fractal distribution of great complexity
and rich structure.
The fractional generalization of integral Maxwell
equations may have applications in the analysis of 
electrodynamical problems involving the fractal stuctures. 
Therefore, it is interesting to numerically solve 
the fractional equations for charged fractals. 
The fractional continuous model can be used to describe dynamics 
and kinetics of the fractal distribution in the plasma physics.  
Extension of this model to describe the dynamical properties
of fractal distribution by fractional generalization 
of magnetohydrodynamics and Vlasov 
equations is currently under study by the author.

%\newpage
%%%%%%%%%%%%%%%%%%%%%
\section{Appendix}

\subsection{Fractional Gauss's theorem}

In order to realize the representation, we derive the fractional 
generalization of Gauss's theorem
\[ \int_{\partial W} ({\bf J}({\bf r},t), d{\bf S}_2) 
=\int_W div( {\bf J}({\bf r},t) ) dV_3 , \]
where the vector ${\bf J}({\bf r},t)=J_k{\bf e}_k$ is a field, 
and $div( {\bf J})={\partial {\bf J}}/{\partial {\bf r}}= 
{\partial J_k}/{\partial x_k}$.
Here and later we mean the sum on the repeated index
$k$ from 1 to 3. Using the relation
\[ d{\bf S}_d=c_2 (d,{\bf r})d{\bf S}_2 , \quad 
c_2(d,{\bf r})= \frac{2^{2-d}}{\Gamma(d/2)} |{\bf r}|^{d-2} , \]
we get
\[ \int_{\partial W} ({\bf J}({\bf r},t),d{\bf S}_d) 
=\int_{\partial W}  c_2(d,{\bf r})  ({\bf J}({\bf r},t) , d{\bf S}_2) . \]
Note that we have $c_2(2,{\bf r})=1$ for the $d=2$. 
Using the usual Gauss's theorem, we get 
\[ \int_{\partial W}  c_2(d,{\bf r}) ({\bf J}({\bf r},t), d{\bf S}_2) =
\int_W  div(c_2(d,{\bf r}) {\bf J}({\bf r},t)) dV_3 . \]
The relation
\[ dV_D=c_3 (D,{\bf r})dV_3 , \quad 
c_3(D,{\bf r})= \frac{2^{3-D} \Gamma(3/2)}{\Gamma(D/2)} |{\bf r}|^{D-3}  \]
in the form $dV_3=c^{-1}_3(D,{\bf r}) dV_D$
allows us to derive the fractional generalization of Gauss's theorem:
\[ \int_{\partial W} ({\bf J}({\bf r},t), d{\bf S}_d)=
\int_W c^{-1}_3(D,{\bf r}) 
div \Bigr( c_2(d,{\bf r}) {\bf J}({\bf r},t) \Bigr) \ dV_D .\]

Analogously, we can get the fractional generalization
of Stokes's theorem in the form
\[ \oint_L ({\bf E},d{\bf l}_{\gamma})=
\int_S  c^{-1}_2(d,{\bf r})
(curl(c_1(\gamma,{\bf r}){\bf E}), d{\bf S}_d) , \]
where 
\[ c_1(\gamma,{\bf r})=
\frac{2^{1-\gamma}\Gamma(1/2)}{\Gamma(\gamma/2)}|{\bf r}|^{\gamma-1} . \]

\subsection{Cantor set and fractional continuous model}

In the paper,
we mention the difference between the real fractal medium structures 
and replacing it by a fractional continuous mathematical model.  
Some quantitative measure of the difference would be helpful.
Note that 
the difference between the real fractal media and fractional 
continuous medium model has an analog of
the difference between the real atomic structure of the media
and the usual continuous medium models of these media.
In order to have some quantitative measure of the applicability
of fractional continuous medium model, 
we can consider the power law for fractal media.
The fractal distribution of charged particles is characterized by 
the law $Q([0,R])\sim R^D$ \cite{Feder}.

%Figures

%%\newpage

\begin{figure}
\centering
\rotatebox{270}{\includegraphics[width=7 cm,height=11 cm]{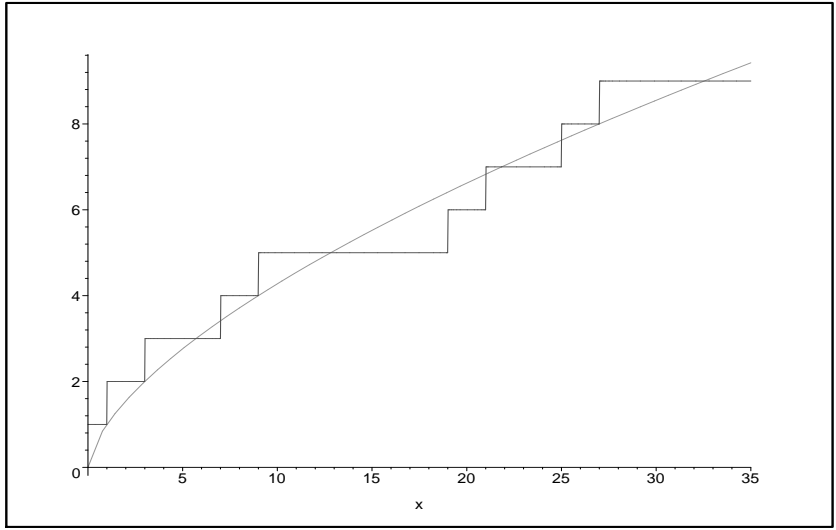}}
\caption{\label{1} 
Charge of fractal distributiom in the interval  $[0;30]$.}
\end{figure}

\begin{figure}
\centering
\rotatebox{270}{\includegraphics[width=7 cm,height=11 cm]{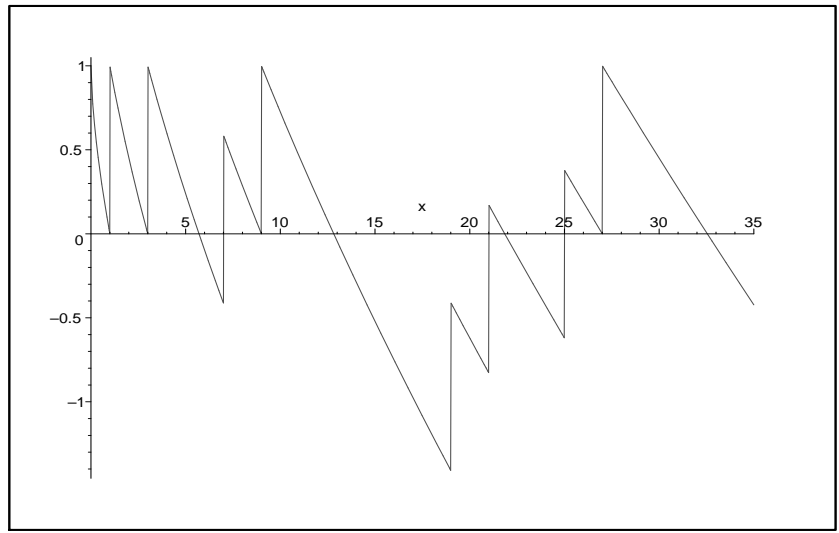}}
\caption{\label{2} 
Difference between the charge of fractal distribution  
and fractional continuous model.}
\end{figure}

The Cantor set is given by taking the interval $[0;x]$, 
removing the open middle third, 
removing the middle third of each of the two remaining pieces, 
and continuing this procedure ad infinitum. 
The Cantor set is sometimes also called no middle third set. 

In Fig. 1, we consider the charge $y=Q([0,x])$
of fractal distribution. 
The fractal distribution is described 
by the Cantor set with fractal dimension $D=\ln(2)/\ln(3)$ \cite{Feder}.
The continuous model for the charge distribution
is described by the continuous line $y=x^D$ 
in the interval $x\in[0,30]$. 

In Fig. 2, we consider the difference between the
charge $y=Q([0,x])$ of fractal distribution 
and the charge that is described by 
fractional continuous distribution  
in the interval $x\in[0,30]$.

\newpage
%%%%%%%%%%%%%%%%%%%%%%%%%%%%%%%%%%%%%%%%%%%%%%%%%%%%%%%%%%%

%%%%%%%%%%%%%%%%%%%%%%%%%%%%%%%%%%%%%%%%%%%%%%
\end{document}